\documentclass[conference]{IEEEtran}
\IEEEoverridecommandlockouts
\usepackage{cite}
\usepackage{amsmath,amssymb,amsfonts}
\usepackage{algorithmic}
\usepackage{graphicx}
\usepackage{textcomp}
\usepackage{xcolor}
\usepackage[utf8]{inputenc}
\def\BibTeX{{\rm B\kern-.05em{\sc i\kern-.025em b}\kern-.08em
    T\kern-.1667em\lower.7ex\hbox{E}\kern-.125emX}}
\usepackage[nopostdot,nogroupskip,nonumberlist]{glossaries}

\newacronym{TPM}{TPM}{tree parity machine}
\newacronym{ANN}{ANN}{artificial neural network}

\begin{document}

\title{Synchronization of Tree Parity Machines using non-binary input vectors}

\author{\IEEEauthorblockN{Miłosz Stypiński}
    \IEEEauthorblockA{
        \textit{AGH University of Science and Technology}\\
        Mickiewicza 30, 30-059\\
        Krakow, Poland \\
        stypinski@agh.edu.pl}
    \and
    \IEEEauthorblockN{Marcin Niemiec}
    \IEEEauthorblockA{
        \textit{AGH University of Science and Technology}\\
        Mickiewicza 30, 30-059\\
        Krakow, Poland \\
        niemiec@agh.edu.pl}
}

\maketitle

\begin{abstract}
    Neural cryptography is the application of \acrlongpl{ANN} in the subject of cryptography. The functionality of this solution is based on \acrlong{TPM}. It uses \acrlongpl{ANN} to perform secure key exchange between network entities. This article proposes improvements to the synchronization of two \acrlongpl{TPM}. The improvement is based on learning artificial neural network using input vectors which have a wider range of values than binary ones. As a result, the duration of the synchronization process is reduced. Therefore, \acrlongpl{TPM} achieve common weights in a shorter time due to the reduction of necessary bit exchanges. This approach improves the security of neural cryptography.
\end{abstract}


\glsreset{TPM}
\glsreset{ANN}

\section{Introduction}
Secure key agreement is one of the basic steps in secure channel establishment. The algorithms responsible for the key ex-change must ensure that no eavesdroppers are able to reproduce the secure key. Applied key agreement protocols are based on mathematical operations which have no computationally efficient inversion, e.g. factorization of large number problem or other derived problems.

Quantum computing poses a real threat to applied cryptography systems. Currently used algorithms, based on public-key cryptography approach, offer conditional security. Efficient derivation of a secure key from exchanged fragmentary information may break the security of the key agreement protocol. Currently, there is one known algorithm - Shor’s algorithm - capable of factorizing large numbers. Hence, it can extract exchanged keys and break all applied asymmetric cipher cryptography  \cite{shor}. However, a successful implementation of this algorithm requires a quantum computer with the sufficient number of qubits.

Some modern cryptography techniques - such as quantum cryptography and neural cryptography – are able to overcome this problem and provide a variety of quantum-proof algorithms. The \gls{TPM} is one such solution. It achieves a key agreement functionality by mutual learning of two artificial neural networks. This paper introduces an accelerated key exchange process of two \glspl{TPM} by utilizing non-binary vectors at the input.

The paper is structured as follows. Section~\ref{section_tpm} presents the architecture of \glspl{TPM}, the process of mutual learning, secure key agreement protocol, exchanged key length and security of \glspl{TPM}. Section~\ref{section_entropy} describes entropy and its appliance in terms of quality assessment of exchanged key. 
Section~\ref{section_methodology} explains the methodology of the performed simulations. Section \ref{section_results} presents an analysis of the gathered results.

\section{Tree Parity Machine}\label{section_tpm}
\Glspl{ANN} are increasingly popular, finding application in fields including security. In \cite{TPM_FIRST} the authors introduce a novel approach for the key agreement functionality implemented with neural networks. Such an approach can be also used for error correction in quantum cryptography systems \cite{NIEMIEC_QUANTUM}.

\subsection{\Acrlong{TPM} architecture}

A \acrlong{TPM} is a two-layered perceptron-structured artificial neural network with discrete weights, binary input and binary output \cite{PERCEPTRON_BOOK}. The input vector $X=[x_{11}, x_{12}, \dots , x_{1n}, \dots , x_{k1}, \dots , x_{kn}],\ K, N, k, n \in \mathbb{N}\ \wedge\ k\leq K\ \wedge\ n \leq N$ has $KN$ elements, where $K$ denotes the number of inputs for each neuron in the first layer, and $N$ indicates the number of neurons in the first layer. Every element $x_{kn}$ of input vector $X$ can have one of two possible values, either $-1$ or $1$.

The first layer consists of neurons similar to the McCulloch-Pitts model \cite{McCulloch1943}. Every input $x_{kn}$ is connected to the $k$-th neuron and has its corresponding weight. The values of the weights are the only difference from the former model. Every weight $w_{ij}$ can take a value between $-L$ and $L$, where $L \in \mathbb{Z}$ is the parameter of the \gls{TPM} and denotes the minimum/maximum possible weight value of the input neurons.

The output of the aforementioned neurons is based on the slightly changed signum function $\sigma$. The formula of the function is presented in \eqref{sign_activation}. It differs from thr regular signum function in that it never returns zero. The value of $0$ is mapped either to $1$ or $-1$, based on whether the side is the sender or the recipient of the communication \cite{REKEYING_ARCH}. The recipient and sender side is denoted by $r$ and $s$, respectively. The parties decide beforehand which side is the sender and the recipient.
\begin{equation}
  \sigma(x^{r/s})=\left\{
  \begin{array}{ll}
    \hphantom{-}1,\ x^r \geqslant 0 \vee x^s > 0 \\
    -1,\ x^r < 0 \vee x^s \leqslant 0
  \end{array}
  \right.
  \label{sign_activation}
\end{equation}

The argument for the neuron's activation function is the sum of the products of the input vector's elements with corresponding weight. The exact formula is presented in \eqref{eq_activation}.
\begin{equation}
  y_k = \sigma (\sum_{n=1}^{N}x_{kn} \cdot w_{kn} ) \label{eq_activation}
\end{equation}

\begin{figure}[htbp]
  \centerline{
    \includegraphics[width=\columnwidth]{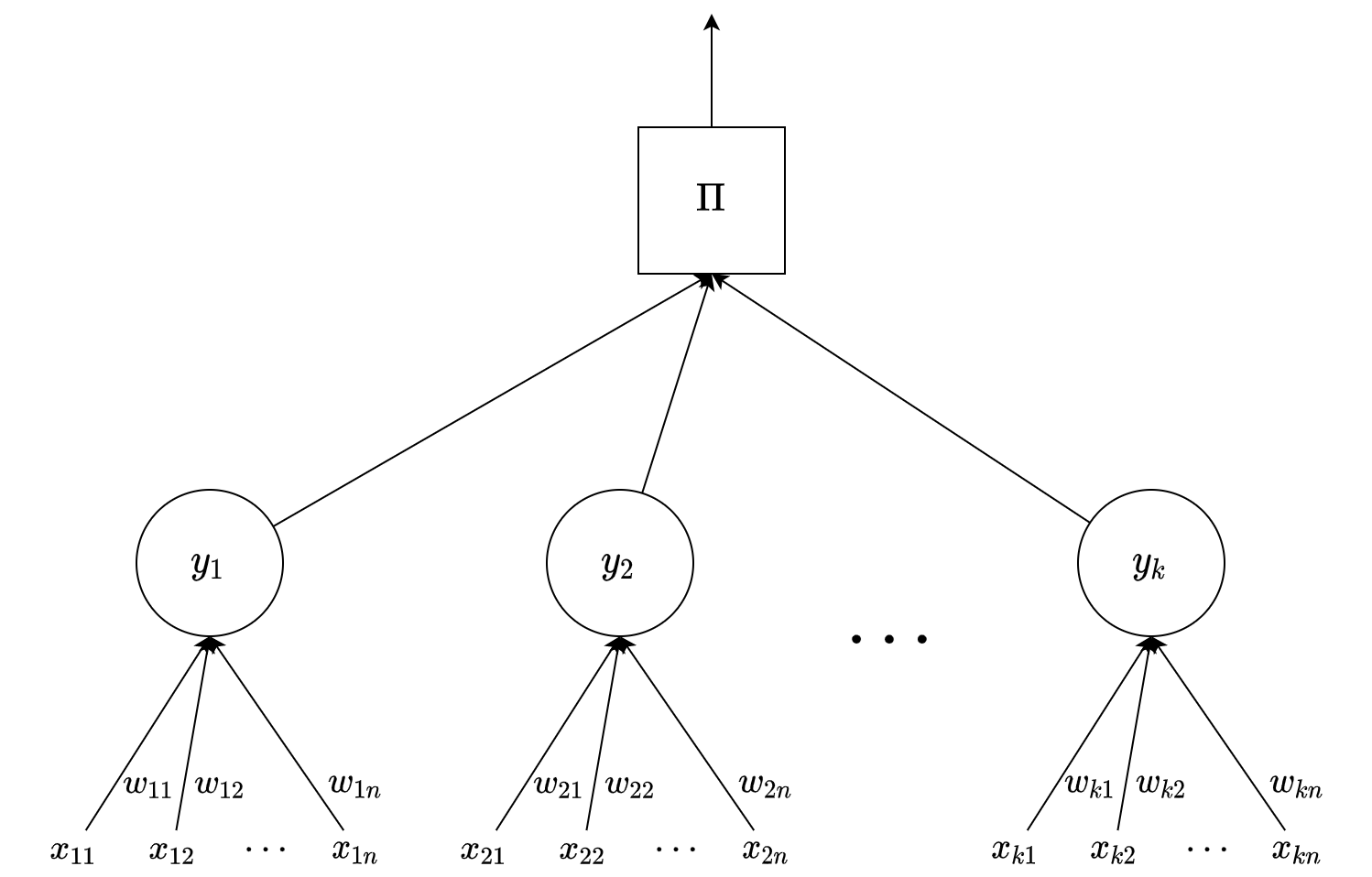}
  }
  \caption{Architecture of the \acrlong{TPM}}
  \label{tpm_architecture}
\end{figure}

The final result $O$ of the \gls{TPM} is the product of each of the outputs from hidden neurons from the first layer \eqref{eq_tpm_final}.
\begin{equation}
  O^{r/s} = \prod_{k}^{K} y_k^{r/s}
  \label{eq_tpm_final}
\end{equation}
The overall architecture of the \gls{TPM} is shown in Figure \ref{tpm_architecture}.

\subsection{Key agreement protocol}

The parties performing the key agreement execute the protocol which results in the secure shared key known only to the participating parties. This is usually achieved through the exchange of some information through an unsecured channel and by performing mathematical operations whose results are only known to the authorized parties~\cite{KeyAgreement}. The first and most popular key agreement protocol was proposed by Diffie and Hellman \cite{DH_Prot}.

\Gls{TPM} offers functionality which can be adopted for key exchange purposes. The protocol for two parties consists of the following steps \cite{REKEYING_ARCH}:
\begin{enumerate}
  \item both participants must agree on all the parameters for the \gls{TPM} $(K, L, N)$ and initialize them with random weights;
  \item \label{proto_step_1} the key agreement participants publicly exchange a previously chosen random input vector $X$;
  \item each party computes the output from their \gls{TPM} and publishes the results;
  \item \label{proto_step_end} if the outputs match, both participants apply the appropriate learning rule which updates the weights of \gls{TPM} accordingly;
  \item steps \ref{proto_step_1}-\ref{proto_step_end} are repeated until full synchronization of both \glspl{TPM} is achieved.
\end{enumerate}
The full synchronization is equivalent to every corresponding weight of both \glspl{TPM} being equal to each other, at which point both \glspl{TPM} are the same.

The aforementioned learning rules are responsible for updating the weights of each \gls{TPM} in such a way that the synchronization process finishes in finite time \cite{FINITE}. There are three different learning rules which can be used in the process of updating weights\cite{TPM_PHD}:
\begin{itemize}
  \item	\textit{Hebbian learning rule}
        \begin{equation}
          w_{kn}(t+1) = w_{kn}(t)+O(t)x_{kn}(t)\Theta(y_k(t), O(t))
          \label{eq_hebbian_lr}
        \end{equation}
  \item	\textit{Anti-Hebbian learning rule}
        \begin{equation}
          w_{kn}(t+1) = w_{kn}(t)-O(t)x_{kn}(t)\Theta(y_k(t), O(t))
          \label{eq_antihebbian_lr}
        \end{equation}
  \item	\textit{Random walk learning rule,}
        \begin{equation}
          w_{kn}(t+1) = w_{kn}(t)+x_{kn}(t)\Theta(y_k(t), O(t))
          \label{eq_random_walk_lr}
        \end{equation}
\end{itemize}
where $\Theta(a,b)$ denotes the function returning $1$ if $a=b$ and $0$ otherwise, and parameter $t$ denotes the iteration in the key agreement algorithm.

The synchronization process of two parity machines is not an deterministic algorithm. The number of iterations is not fixed and depends on the size and parameters of the TPM. However, it is shown that the time is finite and can be easily estimated by users \cite{synchr}. The process takes longer for larger \gls{TPM} sizes ($K$ and $N$) and maximum weight value ($L$). Other factors which affect the number of iterations required for two \glspl{TPM} to finish mutual learning include distribution of initial weights and learning rule \cite{TPM_W_DISTR}.

\subsection{Security of Tree Parity Machines}

Security of key agreement protocol is crucial for communication. Any eavesdropper being able to reproduce the key based on the messages exchanged between parties or any other source breaks the security of the channel. Subsequently, such a situation depreciate the secure key exchange protocol. Hence, it is crucial to assess security of any novel algorithm or protocol.

\Glspl{TPM} have been studied extensively. In \cite{TPM_BF} and \cite{TPM_AES} the authors identify four distinct types of attacks that \gls{TPM} may be vulnerable to:
\begin{itemize}
  \item brute force attack -- research shows that it is impossible to find the exact key as a result of a brute force attack against \glspl{TPM} in polynomial time;
  \item genetic algorithm for weight prediction: it has been shown that only \glspl{TPM} with a single neuron in the second layer are vulnerable to this type of attack;
  \item an-min-the-middle interception attack -- studies show that on average $60\%$ of weights were synchronized in the eavesdropper's \gls{TPM};
  \item sign of weight classification using neural networks -- in \cite{TPM_AES} authors demonstrate that classification using artificial neural networks has near $100\%$ accuracy in determining the sign of the weight in the \gls{TPM}, which reducing the time needed by the brute force attack by almost half.
\end{itemize}
The studies show that, by utilizing these attack vectors, it is possible to gain some information about the key. Hence, cryptosystems should be aware of this threat and counteract it in order to minimize the likelihood of key reconstruction.

\subsection{Man-in-the-middle attack}

Synchronization of two \glspl{TPM} without additional layers of security is a process prone to man-in-the-middle attacks. This attack relies on the possibility of placing a node $C$ between parties $A$ and $B$ performing a key agreement. The node eavesdrops on all the messages shared between $A$ and $B$. Based on information collected, node $C$ may be able to gain unauthorized access to information sent between $A$ and $B$. Moreover, if the nodes are not mutually authenticated, the adversarial party may be able alter the messages accordingly to attempt an attack with a higher probability of success. 

In terms of \glspl{TPM}, man-in-the-middle attacks come down to capturing all the input vectors $X$ and outputs of parties being intercepted. An adversarial \gls{TPM} performs the learning process on acquired data. There are three scenarios to be considered while intercepting the key exchange. Let $A$, $B$ be the parties wishing to exchange the key and let $C$ be an intruder able to perform a man-in-the-middle attack.
\begin{itemize}
    \item If $\Pi_A \neq \Pi_B$ -- no \glspl{TPM} are synchronized during this step.
    \item If $\Pi_A = \Pi_B \neq \Pi_C$ -- only \glspl{TPM} $A$ and $B$ are synchronized, while \gls{TPM} $C$ (attacker) does not update its weights.
    \item If $\Pi_A = \Pi_B = \Pi_C$ -- all the \glspl{TPM} update their weights accordingly.
\end{itemize}
The last scenario brings the adversarial party closer to obtaining the exchanged key. Hence, this situation should be avoided at all costs.

\section{Entropy}\label{section_entropy}
The quality of random numbers generation has a significant impact on the final security of the cryptosystem. A true random number generator produces every available output with equal probability. Unfortunately, computers are incapable of generating fully random numbers. Frequently, numbers are generated based on a pseudo-random number generator. This requires a seed supplied beforehand which is the starting point of the pseudo-random number sequence, and each further number depends on it. Many contemporary implementations lack important features like good mathematical foundations, lack of predictability and cryptographic security \cite{RNG_REPORT}.

Entropy is one of the measures which assesses the quality of the generated numbers. Let us assume the random source generates $I$ different numbers $\alpha_1, \alpha_2, \dots, \alpha_i$ with corresponding probabilities $p_1, p_2, \dots, p_i$. Entropy for such a defined source is presented in \eqref{entropy_eq} \cite{ENTROPY_BASE}.
\begin{equation}
    H(p_1, p_2, \dots, p_i) = -\sum_{i=1}^{I}p_ilog_jp_i
    \label{entropy_eq}
\end{equation}
The base of logarithm $j$ denotes the units in which entropy is measured, e.g. for $2$ and $e$ units are bits and nats respectively \cite{INFTHEORY}. 

\begin{figure}[htbp]
    \centerline{
      \includegraphics[scale=0.28]{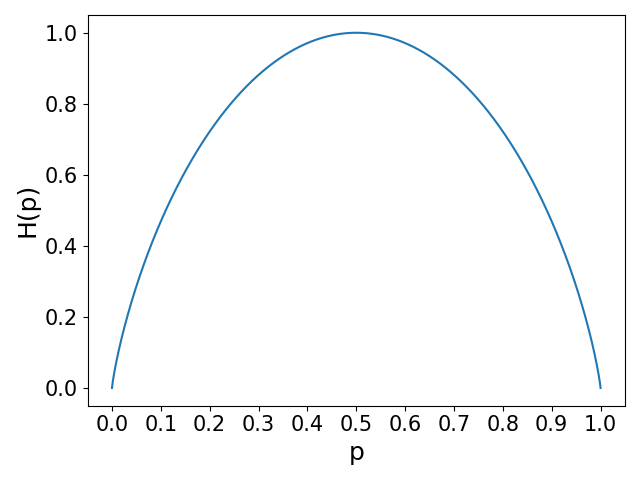}
    }
    \caption{Entropy of the source generating two different values with the same probability}
    \label{entropy_0_1_plot}
\end{figure}

Let us consider a random source which produces two outputs with, either $0$ or $1$ with corresponding probabilities $P(X=0) = p$ and $P(X=1) = 1-p$. The entropy for the described source is presented in \eqref{entropy_0_1_eq}\cite{INFTHEORY}.
\begin{equation}
    H(p) = -plog_jp-(1-p)log_j(1-p)
    \label{entropy_0_1_eq}
\end{equation}
Figure \ref{entropy_0_1_plot} shows the plot of the entropy of the aforementioned two-value random source. The maximum of the function is reached for $p=0.5$ where $H(p) = 1$ which is the equal probability for values $0$ and $1$. Hence, entropy values increase as the probability distribution of $X$ gets closer to the uniform distribution. This can be generalized for sources producing more outcomes.

The entropy function can be used later to assess the quality of the keys generated by different types of \gls{TPM}. Taking into account \eqref{eq_key_length_new} the effective length of a key should depend on the entropy of the synchronized weights (not just their values).

\section{Non-binary input vectors}\label{section_methodology}

The \glspl{TPM} uses binary vectors $X$ for input~\cite{TPM_FIRST} during the synchronization process. There are other approaches presented in \cite{cv_tpm, vv_tpm, cm_tpm} which use complex-valued, vector-valued and chaos generated input vectors accordingly to improve the learning process. Additionally, in \cite{whale_tpm} authors propose whale optimization-based synchronization which results in reduction of the learning process duration. 

This paper introduces a new approach: non-binary input vectors used to synchronize \glspl{TPM} for a secure key agreement protocol. The authors propose that the mutual learning process which uses the vectors with a greater range of possible values of every element influence the synchronization time of two \glspl{TPM}. Simulations performed in the next section verify this proposition and indicate that this approach can significantly increase the security of neural cryptography.

\subsection{Non-binary vector \acrlong{TPM} architecture}

So far, the exact \gls{TPM} was defined by parameters $K, L, N$. In this paper the authors introduce a new parameter $M$, denoting the minimum/maximum value of each element of input vector $X$. Hence, the input vector will have the following form: $X=[x_{11}, x_{12}, \dots , x_{1n}, \dots , x_{k1}, \dots , x_{kn}]$, where $x_{kn} \in \{x: x \in \mathbb{Z} \wedge -M \leq x \leq -1 \vee 1 \leq x \leq M\}$.
Thus, during the synchronization process the entities can use non-binary input vectors, instead of binary vectors which are currently used in practical implementations.

Introducing the $M$ parameter does not affect the architecture of the \gls{TPM} or the learning process. The formulas shown in Section \ref{section_tpm} are still valid despite more divergent values of the learning vectors. However, simulations presented in Section \ref{section_results} show that as the input vectors are more differentiated, the distribution of settled keys is less similar to the uniform distribution. Therefore, an unbiased estimation of key length is required.

\subsection{Agreed key length}

After the synchronization process both parties share identical keys. The keys are distilled from weights of the \gls{TPM} which are the same after the mutual learning process. The key length depends on the size of the TPM as well as the parameter $L$ which indicates the minimum/maximum value the weights may reach during synchronization. Assuming the ideal uniform distribution of the weights, the key length is equal to $K\cdot~N\cdot~log_2(2L + 1)$.
However, the distribution of the weights differs from the uniform distribution \cite{TPM_PHD}. Hence, the entropy should be used to measure the quality of the key exchanged between the parties. The updated key length is defined as follows:
\begin{equation}
  length_{key} = K \cdot N \cdot E(W)
  \label{eq_key_length_new}
\end{equation}
$E(W)$ indicates an average entropy of the weights. Entropy itself is presented in Section \ref{section_entropy}. 

However, the exact distribution of weights is not known beforehand. Taking this fact into consideration, equation \eqref{eq_key_length_new} should be updated. The estimated effective key length shown in equation \eqref{eq_key_length_new_est} uses the estimated entropy based on the simulation results. Additionally, we propose using the floor function in the equation since the unit of effective key are bits.
\begin{equation}
  \hat{length}_{key} = \lfloor K \cdot N \cdot \hat{E}(W) \rfloor
  \label{eq_key_length_new_est}
\end{equation}

It should be noted that equation \eqref{eq_key_length_new} indicates the theoretical maximum key length which can be extracted from mutual weights. However, a dedicated algorithm which equalizes the probability can be used to obtain a cryptography key from an unevenly distributed numerical sequence. This algorithm must be deterministic, since both parties retrieve the cryptographic key from weights simultaneously.

\section{Verification}\label{section_results}
This section presents the impact of the new parameter $M$, indicating the maximum/minimum value of the input vectors during the synchronization process and how it affects the required iterations in the learning process and the quality of the output key.

\subsection{Methodology}
The quality of the output key is measured in its effective length. The effective length is calculated on the basis of \eqref{eq_key_length_new_est}. Further, simulation scenarios cover multiple sets of \glspl{TPM} sizes. For each scenario statistical analysis was prepared based on 1000 simulations. The presented confidence intervals are calculated with a 95\% probability. These scenarios include all possible combinations of parameters $N \in \{40, 50, 60\}$ and $M \in \{1,2,3,4,5\}$. For all simulation scenarios, parameters $K$ and $L$ are equal to $3$ and $5$, respectively. Synchronization time, entropy and effective key length are measured in order to compare the chosen scenarios. Furthermore, we performed man-in-the-middle attack scenarios during which we measured the average synchronization score of the malicious \gls{TPM}.

\subsection{Results}
The synchronization process becomes longer as the size of the \gls{TPM} increases; it also generates a longer key for cryptographic purposes. However, simulations presented in Table \ref{tab_results} reveal that the \gls{TPM} size and parameters are not the only elements that have an impact on the duration of the synchronization process. Multiple simulations were performed with a different values of parameter $M$. An increase of the parameter $M$ value which limits the maximum and minimum possible values $x_i$ of the input vector $X$ reduces the synchronization time significantly. The synchronization time in Table \ref{tab_results} is expressed as a number of output bits exchanged between the parties to achieve full synchronization between the two \glspl{TPM} (learning iterations). Thus, the volume of data exchanged between the parties performing key agreement decreases as the value of parameter $M$ increases.

\begin{table}[htbp]
\caption{Synchronization time of \glspl{TPM} with differentiated input vectors}
\begin{center}
\begin{tabular}{|c|c|c|c|c|c|}
\hline
	&
	&
	\multicolumn{4}{|c|}{\textbf{Synchronization time}} \\
	\cline{3-6} 
	\textbf{$M$} & 
	\textbf{$N$} &
		\textbf{Average}& 
	\textbf{Minimum} & 
	\textbf{Maximum} &
	\textbf{Median} \\
\hline
$ 1 $ & & $ 709\pm490 $ & $ 313 $ & $ 2176 $ & $ 648 $ \\
$ 2 $ & & $ 290\pm216 $ & $ 103 $ & $ 965 $ & $ 273 $ \\
$ 3 $ & $ 40 $ & $ 172\pm138 $ & $ 75 $ & $ 484 $ & $ 156 $ \\
$ 4 $ & & $ 114\pm82 $ & $ 39 $ & $ 314 $ & $ 105 $ \\
$ 5 $ & & $ 84\pm64 $ & $ 28 $ & $ 249 $ & $ 78 $ \\
\hline
$ 1 $ & & $ 714\pm453 $ & $ 333 $ & $ 1981 $ & $ 666 $ \\
$ 2 $ & & $ 316\pm209 $ & $ 145 $ & $ 769 $ & $ 296 $ \\
$ 3 $ & $ 50 $ & $ 172\pm126 $ & $ 69 $ & $ 455 $ & $ 158 $ \\
$ 4 $ & & $ 118\pm88 $ & $ 39 $ & $ 280 $ & $ 108 $ \\
$ 5 $ & & $ 87\pm64 $ & $ 28 $ & $ 216 $ & $ 82 $ \\
\hline
$ 1 $ & & $ 733\pm398 $ & $ 391 $ & $ 1763 $ & $ 715 $ \\
$ 2 $ & & $ 320\pm202 $ & $ 140 $ & $ 652 $ & $ 306 $ \\
$ 3 $ & $ 60 $ & $ 182\pm122 $ & $ 79 $ & $ 421 $ & $ 172 $ \\
$ 4 $ & & $ 121\pm87 $ & $ 43 $ & $ 284 $ & $ 117 $ \\
$ 5 $ & & $ 90\pm70 $ & $ 34 $ & $ 233 $ & $ 81 $ \\
\hline
\end{tabular}
\label{tab_results}
\end{center}
\end{table}

It should be noted that faster synchronization increases security. This is because as the value of parameter M increases, the key agreement process takes less time, hence a longer and more secure key is obtained in a shorter period of time. This makes this solution more competitive among other key exchange protocols.

\subsection{Extrema values effect}
Numerous simulations of the \gls{TPM} learning process using non-binary input vectors led to the discovery of an effect named by the authors as the extrema value effect. Similar effect is shown in \cite{cm_tpm}, however, only binary input vectors are considered in this paper.

\begin{figure*}[htbp]
	\centerline{
	\includegraphics[width=0.93\textwidth]{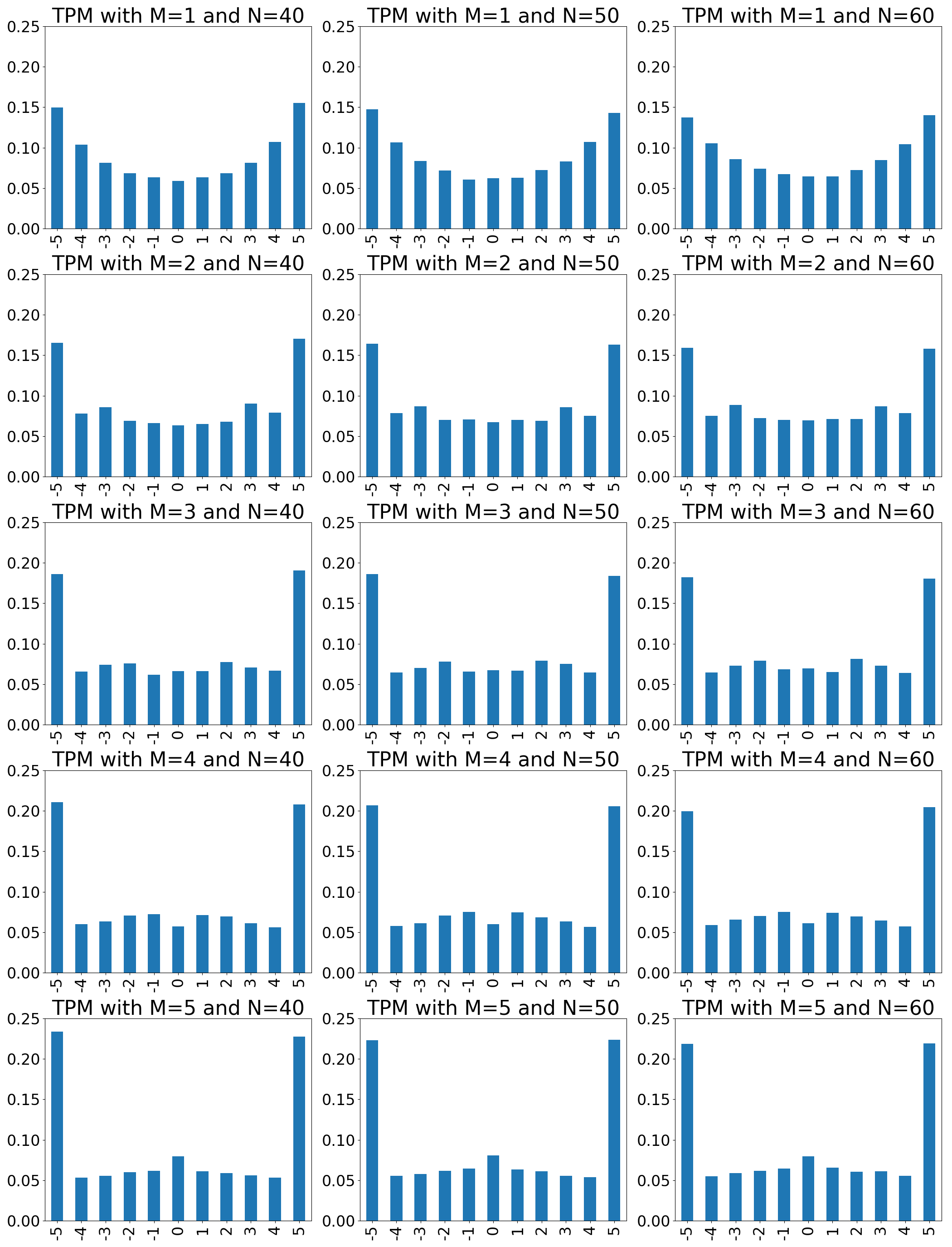}
	}
	\caption{Probability distribution of weights in \glspl{TPM} using non-binary input vectors}
	\label{r_fig1}
\end{figure*}

Faster synchronization times and lower numbers of messages exchanged between users have an impact on the distribution of weights. As the minimum/maximum $x_i$ increases, the probability $P(w_{kn}=M)$ and $P(w_{kn}=-M)$ also increases. As a result, the probability distribution of weights becomes less similar to the uniform distribution. Hence, every weight of the \gls{TPM} carries less random information. The exact distribution of weights is presented in Figure~\ref{r_fig1}.

\begin{figure}[htbp]
    \centerline{
      \includegraphics[scale=0.28]{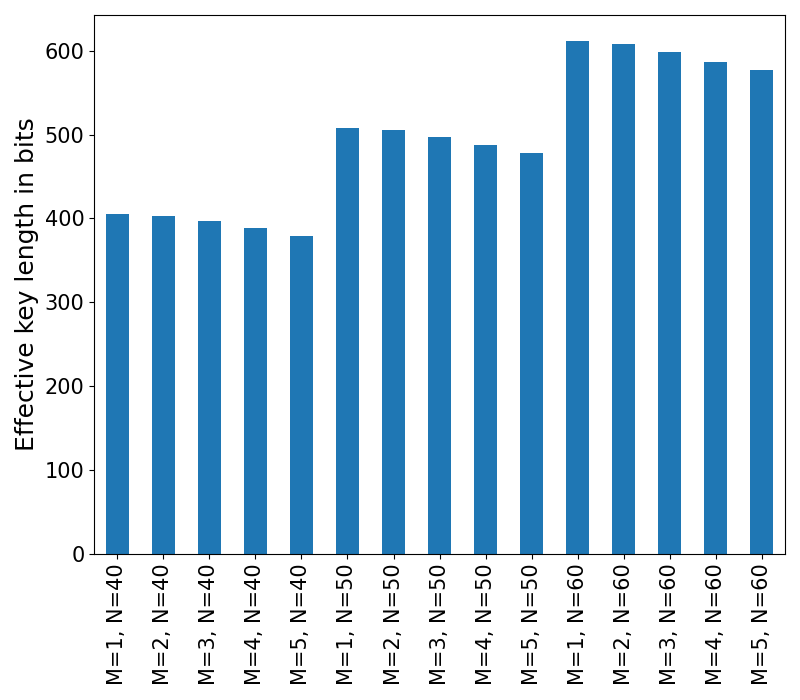}
    }
    \caption{Effective key length of \glspl{TPM} wth different parameters}
    \label{key_length}
\end{figure}

The unequal distribution of weights in the \gls{TPM} results in a reduction of the effective key length since as entropy value becomes lower. Entropy values and effective key lengths are presented in Table \ref{tab_results2}. To visualize the proportion between effective key length, the results for the considered $M$ and $N$ parameters are presented in Figure \ref{key_length}.

\begin{table}[htbp]
	\caption{Entropy and effective key length for different \gls{TPM} parameters}
	\begin{center}
	\begin{tabular}{|c|c|c|c|}
	\hline
		\textbf{$M$} & 
		\textbf{$N$} &
		\textbf{Entropy}& 
		\textbf{Estimated effective key length} \\
		\hline
		$ 1 $ &  & $ 3.374 $ & $ 404 $ \\
		$ 2 $ &  & $ 3.354 $ & $ 402 $ \\
		$ 3 $ & $ 40 $ & $ 3.305 $ & $ 396 $ \\
		$ 4 $ &  & $ 3.238 $ & $ 388 $ \\
		$ 5 $ & & $ 3.158 $ & $ 378 $ \\
		\hline
		$ 1 $ & & $ 3.386 $ & $ 507 $ \\
		$ 2 $ & & $ 3.368 $ & $ 505 $ \\
		$ 3 $ & $ 50 $ & $ 3.315 $ & $ 497 $ \\
		$ 4 $ & & $ 3.248 $ & $ 487 $ \\
		$ 5 $ & & $ 3.186 $ & $ 477 $ \\
		\hline
		$ 1 $ & & $ 3.402 $ & $ 612 $ \\
		$ 2 $ & & $ 3.379 $ & $ 608 $ \\
		$ 3 $ & $ 60 $ & $ 3.325 $ & $ 598 $ \\
		$ 4 $ & & $ 3.263 $ & $ 587 $ \\
		$ 5 $ & & $ 3.204 $ & $ 576$ \\
		\hline
	\end{tabular}
	\label{tab_results2}
	\end{center}
	\end{table}

\begin{table}[htbp]
\caption{Synchronization time of \glspl{TPM} with differentiated input vectors}
\begin{center}
\begin{tabular}{|c|c|c|c|c|c|c|}
\hline
	&
	&
	\multicolumn{4}{|c|}{\textbf{Adversarial \glspl{TPM} synchronization score ($S_{score}$)}} \\
	\cline{3-6} 
	\textbf{$M$} & 
	\textbf{$N$} &
	\textbf{Average}& 
	\textbf{Minimum} & 
	\textbf{Maximum} &
	\textbf{Median} \\
\hline
$1$ & $40$ & $0.174\pm0.047$ & $0.058$ & $0.425$ & $0.167$\\
$2$ & $40$ & $0.18\pm0.055$ & $0.058$ & $0.592$ & $0.167$\\
$3$ & $40$ & $0.205\pm0.079$ & $0.075$ & $0.933$ & $0.183$\\
$4$ & $40$ & $0.243\pm0.101$ & $0.083$ & $0.983$ & $0.208$\\
$5$ & $40$ & $0.28\pm0.136$ & $0.083$ & $1$ & $0.233$\\
\hline
$1$ & $50$ & $0.17\pm0.039$ & $0.073$ & $0.373$ & $0.16$\\
$2$ & $50$ & $0.179\pm0.047$ & $0.08$ & $0.513$ & $0.167$\\
$3$ & $50$ & $0.204\pm0.06$ & $0.087$ & $0.527$ & $0.187$\\
$4$ & $50$ & $0.232\pm0.085$ & $0.087$ & $0.993$ & $0.207$\\
$5$ & $50$ & $0.255\pm0.107$ & $0.067$ & $1$ & $0.213$\\
\hline
$1$ & $60$ & $0.168\pm0.037$ & $0.072$ & $0.361$ & $0.167$ \\
$2$ & $60$ & $0.183\pm0.048$ & $0.061$ & $0.417$ & $0.172$\\
$3$ & $60$ & $0.204\pm0.066$ & $0.067$ & $0.65$ & $0.183$\\
$4$ & $60$ & $0.231\pm0.084$ & $0.078$ & $0.956$ & $0.206$\\
$5$ & $60$ & $0.252\pm0.117$ & $0.089$ & $1$ & $0.211$\\
\hline
\end{tabular}
\label{tab_results3}
\end{center}
\end{table}

\subsection{Susceptibility to a man-in-the-middle attack}
Many research considerations address \gls{TPM} vulnerability to man-in-the-middle attacks. Therefore, simulations with adversarial \glspl{TPM} have been conducted while utilizing learning by non-binary input vectors. 

We assumed the worst-case scenario in which the adversarial neural network was able to eavesdrop on all of the data exchanged between the parties performing the key exchange. During the simulations, the final synchronization score $S_{score}$ was gathered for the adversarial neural network. The synchronization score measures the similarity between two \glspl{TPM}. The more common weights there are, the higher score value is assigned. Hence, the formula needs to return higher values with the progress of the learning process. The formula for calculating the end score is presented in equation \eqref{end_result_eq}. In the following equation, $w^A$ denotes weights of adversarial \gls{TPM} and function $\Theta (a,b)$ is defined in Section \ref{section_tpm}.
\begin{equation}
    S_{score} = \frac{\sum_{k=1}^{K} \sum_{n=1}^{N} \Theta (w_{kn}, w^A_{kn})}{K \times N}
    \label{end_result_eq}
\end{equation}
In terms of security, the attacker's \gls{TPM} should have the lowest synchronization score possible.

\begin{figure}[htbp]
	\centerline{
	\includegraphics[scale=0.35]{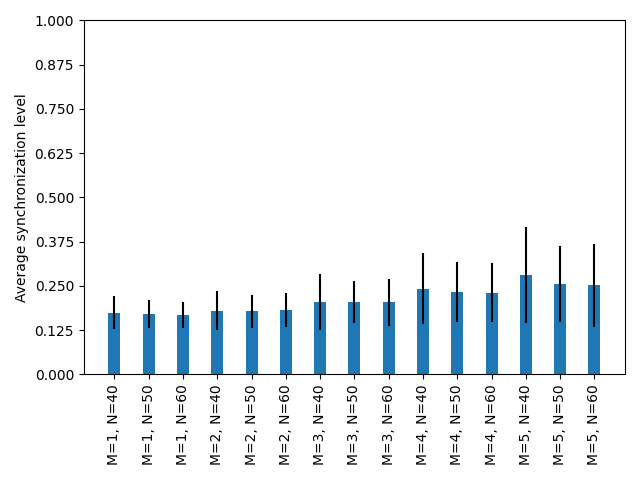}
	}
	\caption{Synchronization score of adversarial \glspl{TPM}}
	\label{syn_plot}
\end{figure}

The synchronization score of adversarial \glspl{TPM} are presented in Table \ref{tab_results3}. Additionally simulation results are shown in Figure \ref{syn_plot} to visualize the relationship between scenarios with different \glspl{TPM}. Increased values of parameter M result in higher median synchronization scores, hence the \gls{TPM} is more prone to man-in-the-middle attacks. When parameter $M$ was equal to $L$, we observed situations where the synchronization score was equal to $1$. This means that the relationship between parameters $M < L$ should be preserved to ensure security. Additionally, the median is inversely proportional to the number of inputs N, therefore the impact of non-binary input vectors on the synchronization score is less clear for larger \glspl{TPM}. Furthermore, the confidence intervals are considerable. This variability makes it difficult to predict the attacker’s malicious \gls{TPM} weights.

\section{Summary}
Correct selection of \gls{TPM} parameters is a key issue of implementing secure key agreement protocols for neural cryptography. It is crucial to find a trade-off between effective key length, synchronization time, and security of the final key which is used by users to protect data in network environment. This comes down to selecting the appropriate network size, extreme values of the weights and learning vectors.

This article proposes an improved way of learning \glspl{TPM}. A significant acceleration of the key agreement process was achieved by utilizing a non-binary input vector. This reduces the volume of data exchanged between the parties performing key agreement. Faster synchronization increases security levels; in particular, it mitigates the risk of the key being obtained by an intruder using a man-in-the-middle attack.
However, the speeding up the process results in an unequal distribution of weights in the \gls{TPM}. This was measured by calculating the effective key length based on the entropy of each weight. The proposed solution was also verified in an insecure environment in which two \glspl{TPM} are a subject to a man-in-the-middle attack.  

We envisage that future work will explore the development of a secure key exchange protocol using non-binary input vectors in \glspl{TPM} during mutual learning. This work will be focused on studying the extrema values effect thoroughly and minimizing the reduction of effective key length.


\section*{Acknowledgement}
This work was supported by the ECHO project which has received funding from the European Union’s Horizon 2020 research and innovation programme under the grant agreement no. 830943.

\bibliography{references} 
\bibliographystyle{ieeetr}
\begin{IEEEbiography}[{\includegraphics[width=1in,height=1.25in,clip,keepaspectratio]{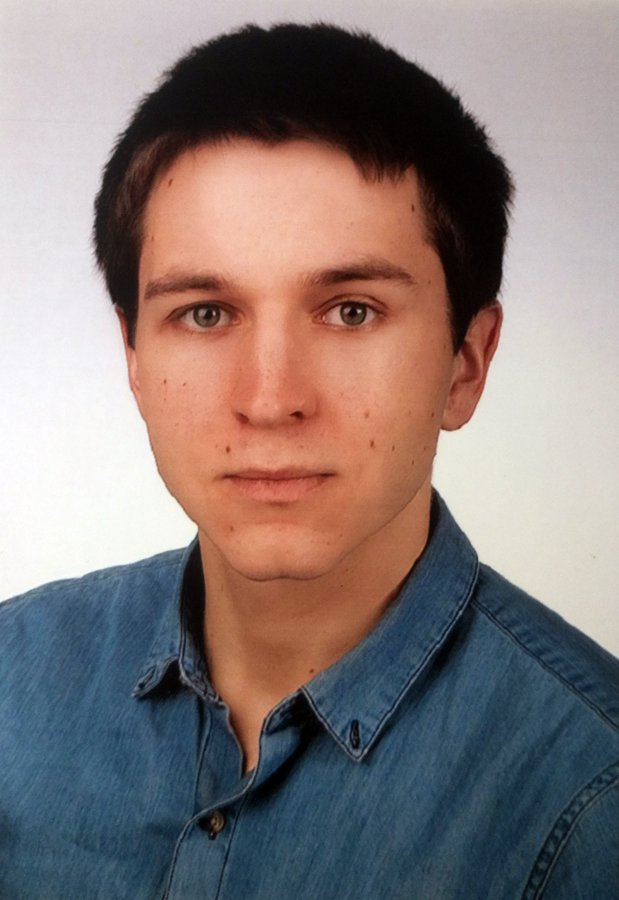}}]{Miłosz Stypiński}received his M.Sc. in ICT from AGH University of Science and Technology, in 2019. Krakow. He is now pursuing a Ph.D. degree in cybersecurity at the same university. His current research interests focus on the application of artificial intelligence algorithms in cybersecurity, in particular symmetric and asymmetric ciphers which use neural networks.
\end{IEEEbiography}

\begin{IEEEbiography}[{\includegraphics[width=1in,height=1.25in,clip,keepaspectratio]{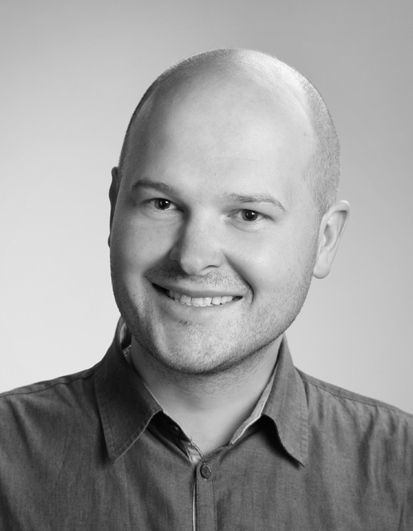}}]{Marcin Niemiec}
is working as a university professor at the Institute of Telecommunications, AGH University of Science and Technology. His research interests focus on cybersecurity and data protection, especially security services, symmetric ciphers, network security, intrusion detection, and quantum cryptography. He has actively participated in 6th and 7th FP European programs (ePhoton/ONE+, BONE, SmoothIT, INDECT), Horizon 2020 Framework Programme (SCISSOR, ECHO), Eureka-Celtic (DESYME), and many national research projects. He co-authored over 90 publications and reports.
\end{IEEEbiography}

\end{document}